# Context Distribution Restoration for Social Surveys: A Recoverability-Adaptive Transport Framework

Lei Zhao[1], Fujin Huang[2], Ling Kang[1], Quan Guo[1]*

[1] Dalian Neusoft University of Information, Dalian, China

[2] No. 760 Research Institute, China State Shipbuilding Corporation Limited (CSSC), Dalian, China

*Corresponding author: guoquan@neusoft.edu.cn*

## Abstract

Social surveys such as CHNS, NHANES, and BRFSS underpin population health and inequality research, yet critical metadata—urban/rural status, gender, and related stratification fields—are often incomplete. External population statistics or survey design information can provide a known prior P(M) over metadata categories. We formalize this setting as Context Distribution Restoration (CDR): recovering sample-level metadata assignments from covariates X while respecting P(M). The core challenge is that metadata recoverability varies by case: some respondents carry strong signals in X, others do not. We define recoverability theoretically as R(M|X)=I(X;M) and approximate it operationally via calibrated predictive uncertainty. We then introduce a recoverability-adaptive transport mechanism within an optimal transport framework to regulate the trade-off between individual evidence and population constraints. Across three large-scale surveys (CHNS, NHANES, BRFSS; up to 67k test samples), we show that unconstrained classifiers (XGBoost) achieve high accuracy but violate P(M) (TV≈0.11), while CDR restores TV<0.001 with minimal accuracy loss. A CHNS case study illustrates interpretable continuum structure. CDR offers a framework for population-consistent metadata restoration in computational social science.

# 1 Introduction

## 1.1 Social problem

National health and nutrition surveys—CHNS (Popkin et al. 2010), NHANES (CDC/NCHS 2023), and BRFSS (CDC 2023)—are foundational data sources for computational social science. Demographic and contextual metadata such as urban/rural residence, gender, and disease status enable subgroup analysis, inequality research, population health monitoring, and social stratification studies (Marmot 2005; Braveman et al. 2011). Yet these stratification fields are frequently incomplete because of privacy suppression, nonresponse, or linkage errors (Little and Rubin 2019). Discarding incomplete cases reduces power and can bias estimates; naive reconstruction that ignores known population margins from census or survey design may distort group shares that downstream analyses treat as population structure.

In practice, analysts often possess a sparse labeled subset for learning predictive relationships and a known prior P(M) over metadata categories from official statistics. Restoring missing metadata therefore has two requirements: (1) individual-level informativeness—assignments should reflect covariates X when signals are present—and (2) preservation of known population structures—reconstructed aggregates should respect P(M). Survey analysis needs both; methods that optimize only individual prediction accuracy address the first requirement incompletely for population-level uses.

## 1.2 Technical gap: prediction accuracy ≠ population-consistent restoration

Individual prediction accuracy does not guarantee population-level validity. Metadata are sometimes easy to recover from covariates X and sometimes hard; a procedure that maximizes per-sample correctness may still under- or over-represent demographic groups relative to P(M). Distorted subgroup proportions may affect prevalence estimates, stratified comparisons, and disparity summaries even when case-level scores look strong. We therefore treat incomplete metadata reconstruction under a known prior as a restoration problem distinct from ordinary classification, and introduce recoverability R(M|X) as the quantity that should govern, per sample, how strongly to follow predictive evidence versus obey the global prior. Context Distribution Restoration (CDR) is a metadata restoration framework for computational social science workflows: given (X, P(M)) and sparse training labels, it produces assignments $\hat{Y}$ that remain informative at the sample level while matching the empirical distribution of $\hat{Y}$ to the known margin c.

CDR differs from related problems in its simultaneous requirements: (i) individual-level informativeness, (ii) population-level consistency with known P(M), and (iii) inference without target-case labels. Standard classification satisfies (i) but not (ii) or (iii); multiple imputation generally focuses on uncertainty-aware recovery of missing variables rather than enforcing exact external population margins as hard constraints; post-hoc Sinkhorn satisfies (ii) but does not incorporate recoverability adaptation.

### 1.3 Contributions

Our contributions are threefold.

**1. Problem formulation.** We formalize missing metadata reconstruction under a known population prior as a distinct problem in computational social science. Unlike classification, which optimizes individual prediction accuracy, CDR requires that predicted assignments respect known population margins while remaining informative at the sample level.

**2. Recoverability-adaptive transport mechanism.** We introduce a mechanism to regulate the trade-off between individual evidence and population constraints within an optimal transport framework. This mechanism provides a way to adapt transport softness based on estimated metadata recoverability.

**3. Empirical demonstration across large-scale surveys.** We validate the framework on three national surveys (CHNS, NHANES, BRFSS) with up to 67k test samples, and provide a CHNS case study illustrating interpretable continuum structure.

## 2 Related work

Classification models maximize individual prediction accuracy. Semi-supervised variants can exploit unlabeled covariates under sparse labels, but neither setting requires predicted subgroup frequencies to match an external prior c. As a result, predicted margins may deviate from known population shares even when accuracy is high—precisely the gap that stratified survey analysis must close.

Missing-data imputation recovers incomplete variables under missingness assumptions (Rubin 1987; Little and Rubin 2019). Imputation improves analytic completeness, yet does not explicitly enforce hard assignment quotas tied to external population constraints P(M). Calibration, raking, and iterative proportional fitting (IPF) adjust population representation by reweighting cases so that weighted margins match known totals; they typically modify sampling weights rather than produce sample-level metadata assignments for each incomplete record.

Optimal transport (Cuturi 2013; Peyré and Cuturi 2019) provides efficient couplings under marginal constraints. Entropy-regularized Sinkhorn assignment can enforce a column prior c at inference and thereby restore distributional margins. Standard Sinkhorn, however, does not by itself incorporate sample-specific recoverability: all rows are softened uniformly unless additional structure is supplied.

CDR does not replace classification, imputation, or calibration/raking/IPF. It combines their complementary concerns in a restoration layer for survey workflows:

individual predictive evidence (costs from covariates and context anchors) plus population distribution constraints (Sinkhorn under P(M)), with a recoverability-adaptive transport mechanism so that easy-to-recover cases follow model costs and hard-to-recover cases defer toward the prior. Relative to prototype-learning pipelines, CDR uses context anchors exclusively to define transport costs C—not as an independent representation-learning objective—while transport adaptation regulates the accuracy–fidelity trade-off under the margin contract.

# 3 Methods

## 3.1 Problem: Context Distribution Restoration (CDR)

Let X ∈ ℝ^{n×d} denote tabular covariates and M ∈ {0,…,K−1}^n a categorical metadata attribute. Official statistics provide a known prior

$$P(M) = c = (c_0, \dots, c_{K-1}), \sum_{k=0}^{K-1} c_k = 1. \quad (1)$$

During training a fraction ρ of labels is retained (ρ=0.1 unless noted); the remainder are masked. At inference the model receives only X and c; held-out labels evaluate accuracy. CDR seeks assignments Ŷ whose distribution matches c while remaining informative about M. The recoverability-adaptive transport objective is

$$\min_T \langle T, C \rangle + \lambda_1 \cdot \mathcal{D}(T1 \| P(M)) + \lambda_2 \cdot \mathcal{L}_{recover}(T, X), \quad (2)$$

where T is a transport plan, C a sample-to-class cost matrix, $\mathcal{D}$ penalizes deviation of column marginals T**1** from P(M), and ℒ_recover encodes recoverability-weighted fidelity to model predictions. $\lambda_1$ and $\lambda_2$ trade population constraint against per-sample recoverability. In our implementation, $\mathcal{D}$ is KL divergence KL(T**1**‖c) for gradient-based optimization and is reported as Total Variation for interpretability.

### 3.2 Recoverability estimation

Recoverability quantifies how much information X carries about M. Theoretically, we define

$$R(M|X) = I(X;M) = H(M) - H(M|X). \quad (3)$$

However, exact mutual information estimation in high-dimensional survey data is challenging. We therefore use an operational approximation based on calibrated predictive uncertainty. For sample i, let p_i be the predicted probability vector. We define

$$\hat{R}_i = 1 - \frac{H(p_i)}{\log K} = 1 - \frac{-\sum_k p_{i,k} \log p_{i,k}}{\log K}, \quad (4)$$

where K is the number of classes. This proxy ranges from 0 (maximum entropy, complete ambiguity) to 1 (deterministic prediction, high recoverability). It is a practical approximation, not an exact estimate of I(X;M). Alternative calibrated uncertainty estimators can be incorporated without changing the CDR framework. High estimated recoverability → stronger adherence to model-based transport costs (trust the prediction); low estimated recoverability → softer, prior-dominated assignment. An optional ambiguity head $\alpha \in [0,1]$ can also serve as a learnable proxy for sample-specific Sinkhorn entropy $\varepsilon_i$.

### 3.3 Context anchors (prototypes)

Context anchors (prototypes) define the cost matrix C for optimal transport. Rather than serving as a standalone representation learner, the prototype bank provides geometrically meaningful reference points: the transport cost between a sample and each anchor determines how the sample will be allocated under the prior constraint. EMA updates ensure anchors remain stable under sparse supervision. Cost-matrix discriminability—not silhouette scores—is the operational role of anchors within CDR.

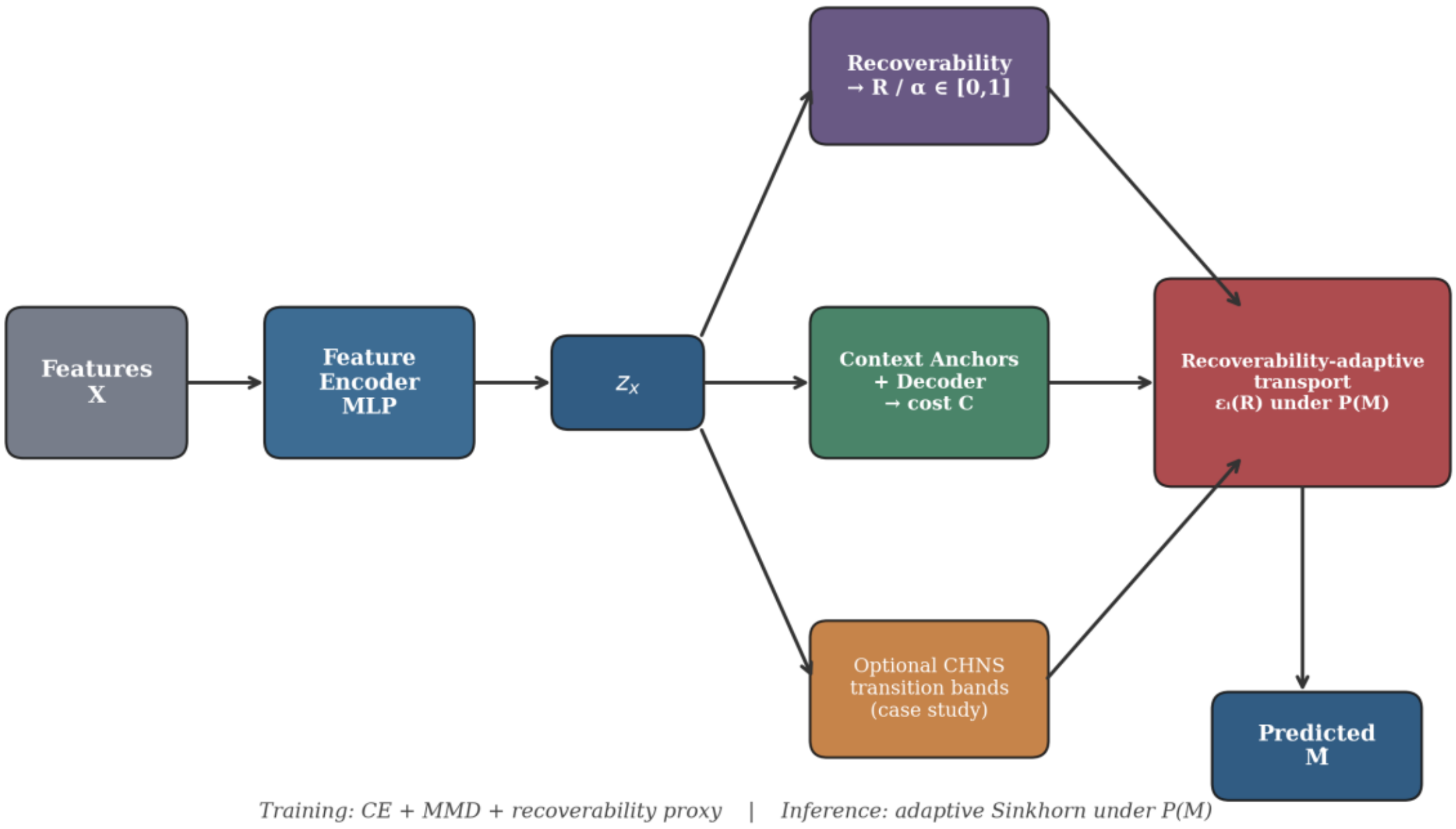

Figure 1 CDR system architecture. Features X are encoded to z_x; context anchors and a decoder produce class costs; recoverability (α) modulates per-sample Sinkhorn entropy under prior P(M).

Encoder. An MLP maps X to latent embeddings:

$$z_x = f_\theta(X), z_x \in \mathbb{R}^{128}. \quad (5)$$

Each class k maintains a prototype anchor p_k updated by exponential moving average:

$$p_k^{(t+1)} = m \cdot p_k^{(t)} + (1-m) \cdot \frac{1}{|S_k|} \sum_{z \in S_k} z, \quad (6)$$

with momentum m=0.99 and S_k labeled embeddings of class k. Cosine similarity to anchors yields costs $C_{ij}$ between sample i and class j for transport.

### 3.4 Recoverability-adaptive transport mechanism

Given cost matrix C and column prior c, entropy-regularized Sinkhorn seeks

$$T^{\star} = \text{\argmin}_{T \in \Pi(r,c)} \langle T, C \rangle - \sum_i \varepsilon_i H(T_{i\cdot}), \quad (7)$$

where row marginal r is uniform 1/n and $\varepsilon_i$ is sample-specific. The OT marginal constraint provides distributional restoration: Sinkhorn enforces the column prior c ≈ P(M). A recoverability-adaptive transport mechanism regulates how much individual predictive evidence contributes under this constraint. Larger $\varepsilon$ softens the coupling toward the maximum-entropy plan under marginals c (prior dominates row-wise allocation); smaller $\varepsilon$ sharpens adherence to costs C (model prediction dominates). The recoverability-adaptive mechanism therefore targets $\varepsilon_i \downarrow$ when R is high and $\varepsilon_i \uparrow$ when R is low, thereby controlling the accuracy–fidelity trade-off—not the margin match itself. A practical interface sets $\varepsilon_i$ from R̂_i or an ambiguity head $\alpha$; under the present $\alpha$ schedule on CHNS, adaptive $\varepsilon$ matches fixed global $\eta$ on accuracy and TV (Table 3b)—we report it as control infrastructure rather than an empirical accuracy upgrade. Hard labels are obtained by quota-preserving rounding of T.

The purpose of recoverability weighting is not to replace the marginal constraint, but to regulate the trade-off between model evidence and prior information. This adjusts the entropic regularization schedule within Sinkhorn transport, rather than replacing the OT objective.

### 3.5 Implementation details

Training retains $\rho$=0.1 of labels (fixed seed), masks the remainder, and optimizes masked cross-entropy plus auxiliary alignment losses (MMD batch alignment; $\alpha$ consistency with GMM posteriors on labeled embeddings). AdamW (lr=$10^{-3}$, weight decay $10^{-4}$), batch size 512, cosine decay, early stopping on validation accuracy. Sinkhorn runs at inference with column constraint c.

For binary tasks with prototype-collapse risk, we optionally apply hyperspherical EMA (Adaptive Prototype Activation, APA); this is an engineering stabilization, not a core algorithmic contribution. APA selects among CONTINUUM / STANDARD / STABILIZED regimes after a short warmup based on a collapse-risk probe.

Graduated transition bands between adjacent anchors are used only in the CHNS urban–rural case study (Section 4.6) to illustrate continuum interpretability; they are not a core CDR algorithmic contribution.

# 4 Results

## 4.1 Datasets

We evaluate three metadata restoration tasks (Table 1). Training retains 10% of labels; test labels are withheld from the model. Only X and P(M) are provided at inference.

**Table 1 Survey datasets and metadata restoration tasks.**

| Dataset | Target | Classes | Feat. | N train | N test | Prior (ex.) |
|---|---|---|---|---|---|---|
| CHNS | Urban/Rural | 2 | 7 | 71,494 | 15,387 | 33.3% urban |
| NHANES | Gender | 2 | 44 | 5,732 | 1,234 | 46.2% |
| BRFSS | Diabetes | 2 | 23 | 314,340 | 67,649 | 15.1% pos. |

## 4.2 Restoration accuracy and distribution fidelity

Table 2 reports Sinkhorn-constrained accuracy (mean±s.d., three seeds) for CDR versus anchor-free CDR—a simplified ablation of the stack—with unconstrained XGBoost as an individual-prediction reference that ignores P(M). Across tasks, OT under the column prior keeps TV<0.001: the marginal constraint is the primary driver of distributional restoration, while the recoverability-adaptive mechanism regulates how strongly individual predictive evidence contributes under that constraint. The largest significant gain over anchor-free CDR is on BRFSS diabetes (+5.27 pp, p=0.010); CHNS and NHANES differences are small. Classical models with the same Sinkhorn readout achieve comparable constrained accuracy (Table 2b), confirming that population fidelity is driven by OT while full CDR integrates adaptive transport costs and context anchors.

**Table 2 Multi-seed Sinkhorn accuracy: CDR versus anchor-free CDR (%, seeds 42/43/44). *Diabetes multi-seed runs use a stratified 60k subset.**

| Dataset | Anchor-free CDR | CDR | Δ (pp) | p | XGB Argmax | TV (CDR) |
|---|---|---|---|---|---|---|
| CHNS Urban/Rural | 69.41±0.55 | 69.28±0.38 | −0.13 | 0.36 | 71.23% | 0.0006 |
| NHANES Gender | 79.12±0.92 | 79.34±0.84 | +0.22 | 0.27 | 81.77% | 0.0008 |
| BRFSS Diabetes* | 76.03±1.03 | 81.30±0.12 | +5.27 | 0.010 | 84.86% | 0.0005 |

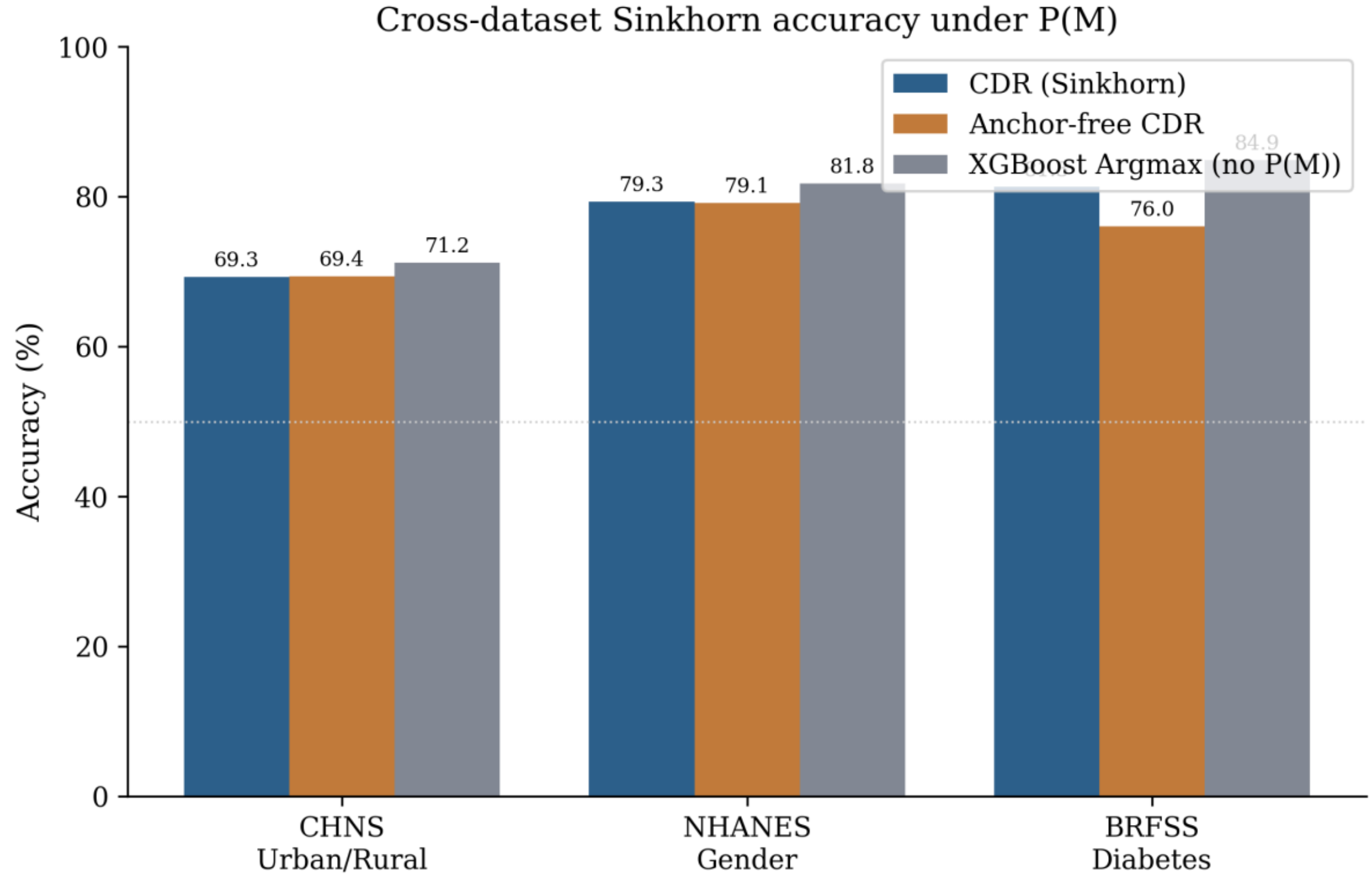

Figure 3 Cross-dataset Sinkhorn accuracy for CDR, anchor-free CDR, XGBoost+Sinkhorn, and LogReg+Sinkhorn under P(M). Dashed segments mark unconstrained XGBoost Argmax (no prior). Hatched bars aid grayscale reading; error bars on CDR are multi-seed s.d. (seeds 42–44).

**Table 2b Classical baselines with Sinkhorn (SK) post-processing under P(M), ρ=0.1.**

| Dataset | LogReg | LogReg+SK | XGB | XGB+SK | CDR+SK |
|---|---|---|---|---|---|
| CHNS Urban/Rural | 66.68% | 68.56% | 71.23% | 69.51% | 68.95% |
| NHANES Gender | 77.96% | 77.96% | 81.77% | 82.66% | 79.50% |
| BRFSS Diabetes | 70.01% | 81.70% | 84.86% | 81.59% | 81.54% |

XGBoost attains strong per-sample accuracy yet can violate P(M): on CHNS the predicted urban share is 22.2% versus 33.3% true (TV≈0.110), while Sinkhorn under the known prior recovers 33.2% (TV≈0.0006)—illustrating why population-consistent restoration differs from unconstrained prediction. Figure 3 visualizes the same Sinkhorn accuracy comparisons across datasets and baselines, with error bars from multi-seed runs (seeds 42–44). These results indicate that marginal restoration is largely achieved by OT constraints, while the recoverability-adaptive mechanism determines how individual-level information is preserved under this constraint.

### 4.2.1 Recoverability-stratified transport analysis

To test whether recoverability modulates transport strength as designed, we compute $\hat{R}_i$ (Section 3.2) for each CHNS test sample and stratify into three groups by terciles: high, medium, and low. For each group we measure (i) the proportion of samples whose Sinkhorn assignment differs from Argmax (assignment change rate) and (ii) the accuracy gap between Sinkhorn and Argmax readouts.

**Table 2e Recoverability-stratified transport analysis on CHNS (tercile cutoffs $\hat{R}$≈0.02/0.29).**

| Recoverability group | N | Assignment change rate | ΔAcc (Argmax→SK, pp) |
|---|---|---|---|
| High | 5129 | 0.0% | +0.00 |
| Medium | 5129 | 2.8% | -0.57 |
| Low | 5129 | 38.0% | -3.24 |

Low-recoverability samples are reassigned more frequently (38.0% vs 0.0% for high-R) and incur a larger Argmax→Sinkhorn accuracy change (-3.24 vs +0.00 pp), consistent with the design: when X is ambiguous, the prior should dominate row-wise allocation under the shared P(M) constraint. This pattern supports recoverability as a modulator of the accuracy–fidelity trade-off; distributional restoration itself is attributable to the OT marginal constraint.

## 4.3 Mask sensitivity

Recoverability depends on supervision density. We vary label retention $\rho \in \{0.1,0.3,0.5,0.7,0.9\}$ on CHNS (equivalently mask rates 90%→10%) and compare Argmax versus Sinkhorn under fixed P(M). Hypothesis: when labels are scarce (low ρ), unconstrained predictions leave a larger fidelity gap and Sinkhorn induces a larger accuracy change; when labels are abundant (high ρ), that accuracy trade-off shrinks—while unconstrained high-capacity models may still violate P(M).

**Table 2c Mask sensitivity on CHNS (XGBoost ± Sinkhorn). ρ = fraction of training labels retained.**

| ρ (labels kept) | Argmax Acc (%) | Sinkhorn Acc (%) | ΔAcc (pp) | Argmax TV | Sinkhorn TV |
|---|---|---|---|---|---|
| 0.1 | 71.23 | 69.51 | -1.72 | 0.1117 | 0.0000 |

| 0.3 | 72.04 | 70.83 | -1.21 | 0.1180 | 0.0000 |
|---|---|---|---|---|---|
| 0.5 | 72.81 | 71.50 | -1.31 | 0.1252 | 0.0000 |
| 0.7 | 73.16 | 71.98 | -1.18 | 0.1256 | 0.0000 |
| 0.9 | 73.24 | 72.08 | -1.16 | 0.1248 | 0.0000 |

At ρ=0.1 (sparse labels), XGBoost Argmax reaches 71.2% with TV=0.112; Sinkhorn restores TV=0.0000 at ΔAcc=-1.72 pp. At ρ=0.9 (dense labels), Argmax accuracy rises to 73.2% and Sinkhorn ΔAcc shrinks to -1.16 pp, yet Argmax TV remains large (0.125). Thus denser labels improve recoverability of individual labels but do not by themselves guarantee population fidelity—the OT marginal constraint under P(M) remains necessary.

**Table 2d Mask sensitivity on CHNS (CDR ± Sinkhorn). Training uses the same early-stopping protocol as Table 2 (max 35 epochs, patience 12; single seed 42). Table 2 reports fully converged multi-seed means. Sinkhorn TV stays ≈0.0006 across ρ.**

| ρ (labels kept) | Argmax Acc (%) | Sinkhorn Acc (%) | ΔAcc (pp) | Argmax TV | Sinkhorn TV |
|---|---|---|---|---|---|
| 0.1 | 70.27 | 68.92 | -1.36 | 0.1256 | 0.0006 |
| 0.3 | 70.26 | 68.94 | -1.32 | 0.1479 | 0.0006 |
| 0.5 | 70.09 | 68.94 | -1.15 | 0.1400 | 0.0006 |
| 0.7 | 70.35 | 68.89 | -1.46 | 0.1532 | 0.0006 |
| 0.9 | 70.37 | 68.93 | -1.44 | 0.1557 | 0.0006 |

Under CDR, Sinkhorn Acc stays near the Table 2 operating point (68.9%–68.9%) while Sinkhorn TV remains ≈0.0006. Argmax TV is higher at low ρ, consistent with weaker recoverability under sparse labels.

### 4.4 Distribution fidelity

Table 3 and Figure 4 compare Argmax versus Sinkhorn readouts of the same trained models. Argmax yields TV in the 0.01–0.13 range; Sinkhorn reduces TV below 0.002 with modest accuracy changes—demonstrating that the OT marginal constraint, not the base predictor alone, restores P(M). Reducing TV is not only a machine-learning metric: population-level distortion in reconstructed metadata may affect prevalence estimates, subgroup comparisons, and disparity analysis in computational social science workflows. We do not claim demonstrated downstream improvement here; the point is that unconstrained reconstruction can leave margins that are inconsistent with known population structure.

**Table 3 Argmax versus Sinkhorn readout (TV to P(M)).**

| Dataset | Argmax TV | Sinkhorn TV | TV ratio |
|---|---|---|---|
| CHNS | 0.134 | 0.0006 | 223× |
| NHANES Gender | 0.012 | 0.0008 | 15× |
| BRFSS Diabetes | 0.1266 | 0.0007 | 181× |

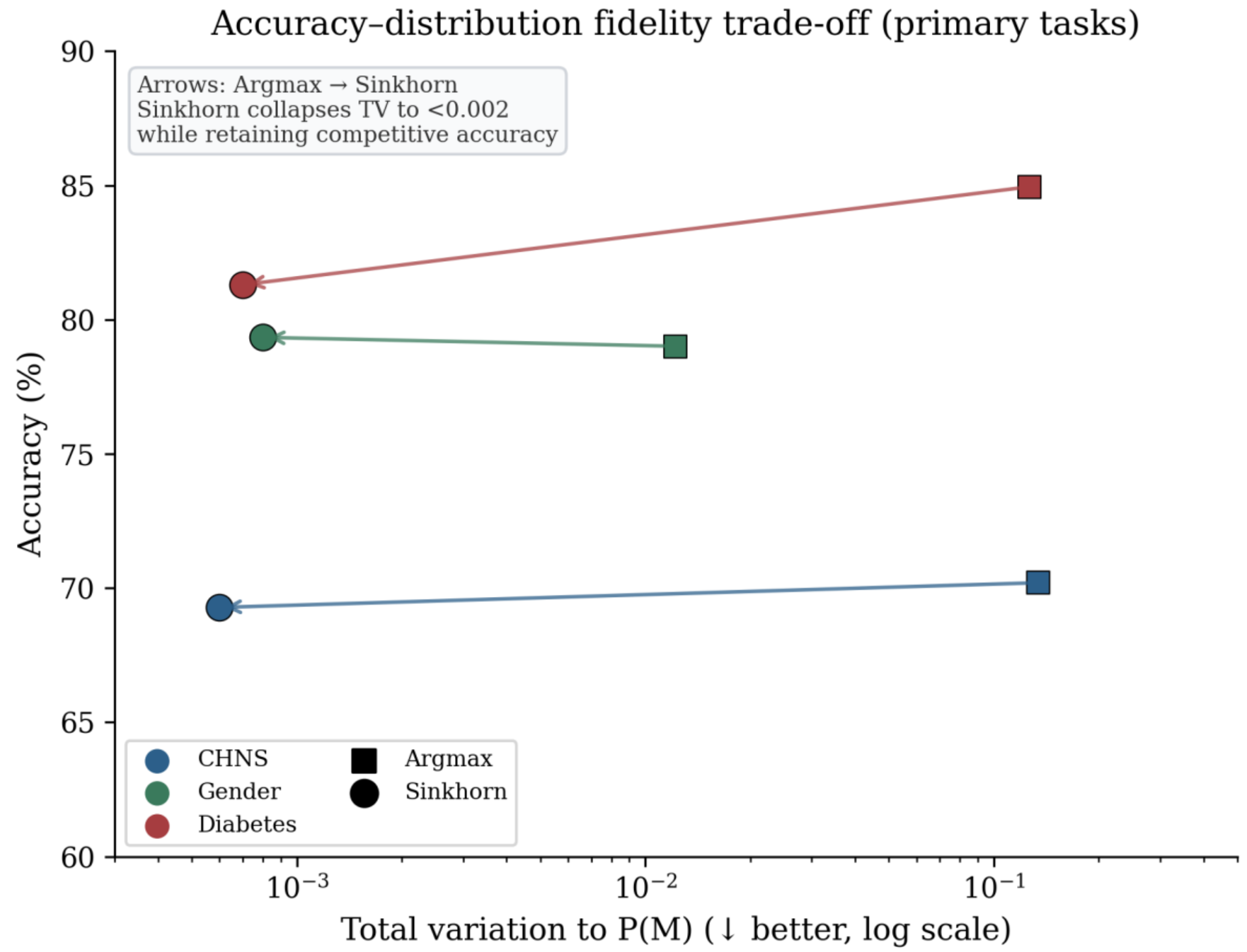

Figure 4 Accuracy–distribution fidelity trade-off. Squares: Argmax; circles: Sinkhorn under P(M). Lower TV indicates closer match to known population margins relevant for stratified survey analysis.

### 4.4.1 ε sensitivity (implementation detail)

On a fixed CHNS model, global and sample-specific ε schedules achieve essentially identical Sinkhorn accuracy (≈69.1%) and TV≈0.0004 (Table 3b). Under the present α estimates, recoverability-modulated ε matches fixed η on both accuracy and distributional fidelity—consistent with Sinkhorn's marginal constraint as the primary restorer of P(M), while ε adaptation remains available to regulate the accuracy–fidelity trade-off under stronger external uncertainty signals.

**Table 3b CHNS Sinkhorn entropy schedule sensitivity.**

| Schedule | Accuracy | TV |
|---|---|---|
| Argmax (ablation) | 70.27% | 0.185 |
| Threshold hybrid | 70.14% | 0.171 |
| Global $\eta$=0.1 (fixed) | 69.14% | 0.0004 |
| Sample-specific $\varepsilon_i(\alpha)$ | 69.10% | 0.0004 |
| $\varepsilon \propto (1+\text{ambiguity})$ | 69.09% | 0.0004 |
| $\varepsilon \propto \alpha(1-\alpha)$ | 69.08% | 0.0004 |

## 4.5 Ablation

Table 4 ablates CDR components on CHNS. Removing Sinkhorn (Argmax readout) inflates TV consistent with Table 3, confirming that the OT/Sinkhorn constraint is the primary driver of distribution fidelity. Other removals change hard accuracy modestly. The adaptive mechanism for ε regulates the accuracy–fidelity trade-off under that constraint rather than substituting for it.

**Table 4 Component ablation on CHNS urban/rural prediction.**

| Config | Removed | Hard Acc. | Δ vs Full |
|---|---|---|---|
| A | None (Full) | 68.89% | — |
| B | Transition bands | 69.25% | +0.36% |
| C | MMD | 69.10% | +0.21% |
| D | Ambiguity head α | 69.05% | +0.16% |
| E | Sinkhorn (Argmax) | 68.79% | −0.10% |
| F | Inference-only Sinkhorn | 68.85% | −0.04% |
| H | Anchor-free CDR | 68.85% | −0.04% |

## 4.6 Case study: CHNS urban–rural continuum

On CHNS, feature-space overlap (Bayes error 4.9%, overlap 4.1%) activates the optional continuum gate (Table 5). In prototype space, 939/15,387 test cases (6.1%) fall in the boundary band and concentrate along the Urban↔Rural interpolant: 79.9% of mid-axis cases carry transition labels versus a 6.1% base rate (Figure 5). This is an interpretability case study illustrating the framework's capacity to expose continuum structure; it is not an accuracy claim (Table 4 Config B). Geographic validation of this band as peri-urban remains future work. APA stabilization metrics appear in Supplementary Table S1.

**Table 5 Label-free overlap gate for continuum bands (CHNS case study).**

| Dataset | Feat. Bayes | Feat. overlap | Decision | n_levels |
|---|---|---|---|---|
| CHNS | 4.9% | 4.1% | Use | 1 |
| NHANES Gender | 0.9% | 0.7% | Skip | 0 |
| BRFSS Diabetes | <0.05% | <0.05% | Skip | 0 |

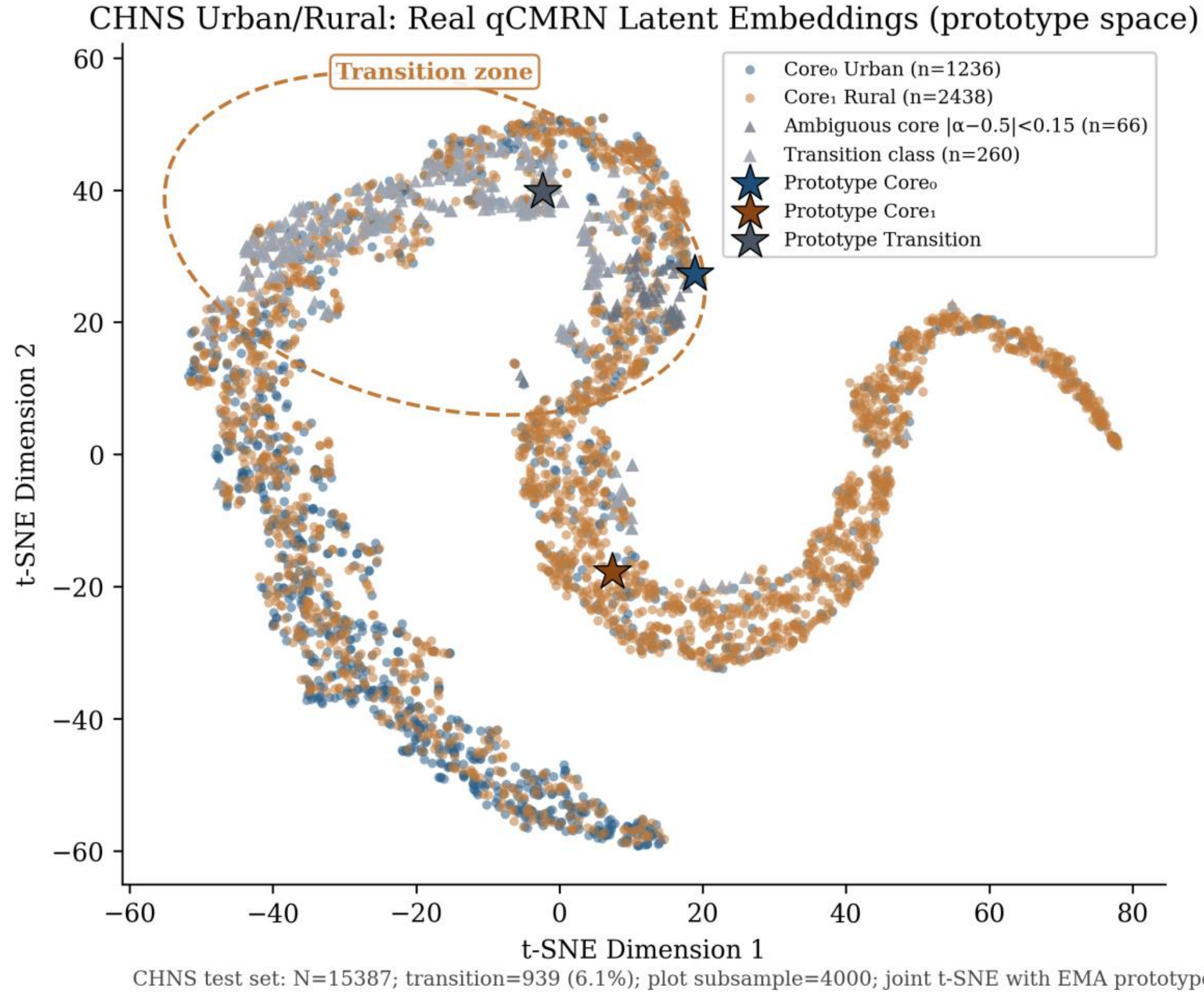

Figure 5 CHNS prototype-space t-SNE (n=4,000 of 15,387 test embeddings). Stars: urban, rural, and transition anchors; triangles: boundary-band cases (6.1%).

## Supplementary Material

### Supplementary Table S1 APA prototype stabilization

For binary tasks with collapse risk we optionally apply hyperspherical EMA (Adaptive Prototype Activation). This is an engineering stabilization of context anchors, not a core algorithmic contribution, and does not alter Sinkhorn marginals. APA raises BRFSS silhouette from −0.056 to 0.201.

**Supplementary Table S1 APA regime selection and prototype-space metrics (seed 42).**

| Dataset | APA mode | Silhouette | DBI ↓ |
|---|---|---|---|
| CHNS | CONTINUUM | 0.229 | 2.64 |
| NHANES | STANDARD | 0.100 | 4.04 |
| BRFSS | STABILIZED | 0.201 | 2.27 |

# 5 Discussion

## 5.1 Metadata restoration versus unconstrained prediction

CDR addresses how to restore incomplete survey metadata when analysts have covariates X and a known prior P(M). The OT/Sinkhorn marginal constraint restores distributional fidelity (e.g., CHNS unconstrained TV≈0.11 reduced to TV<0.001 under Sinkhorn; Tables 2–3), while the recoverability-adaptive mechanism regulates how strongly individual predictive evidence contributes under that constraint. When ρ is low or covariates are weak, the prior should dominate ambiguous cases; when estimated recoverability is high, model costs should govern transport. Individual-level predictive accuracy alone does not ensure population-consistent reconstruction—a distinction that matters for survey uses that depend on population aggregates.

## 5.2 Implications for computational social science

CDR is designed as a preprocessing and restoration layer before downstream analyses, not as a substitute for substantive models of inequality, health, or stratification. By producing sample-level metadata assignments whose aggregates respect known population margins, the framework provides a principled bridge between predictive scoring of incomplete cases and population-level validity. Potential applications include demographic stratification, health inequality studies, and population statistics that rely on reconstructed contextual fields. We do not claim that CDR improves causal inference; whether restored labels alter particular disparity indices or policy conclusions remains an empirical question for domain-specific re-analyses.

### 5.3 Transition bands (demoted case study)

Graduated transition bands on CHNS illustrate how context anchors expose continuum structure between urban and rural cores; geographic validation against peri-urban administrative labels remains future work. Transition is an interpretability aid in the case study, not a core CDR contribution or accuracy objective.

### 5.4 Limitations

First, CDR is evaluated on three binary metadata restoration tasks. The framework is not restricted to binary metadata; extension to richer multi-category contextual fields represents an important future direction. Feature sets in the present study are comparatively compact. Second, population priors may be uncertain: official margins can be dated, smoothed, or estimated with error. Extensions to uncertain priors and probabilistic marginal constraints are natural directions for future work. Third, recoverability estimation depends on uncertainty calibration; predictive-entropy proxies are related to, but not identical with, mutual information $I(X;M)$, and their quality tracks model calibration (Section 3.2). Empirically, recoverability-modulated $\varepsilon$ does not yet outperform fixed $\varepsilon$ on CHNS under the present $\alpha$ estimates (Table 3b). APA and transition bands remain non-core engineering and case-study tools. End-to-end re-analysis of published disparity indices under constrained labels is left to future work. Sinkhorn adds modest inference cost (≈26 ms per 10k samples on BRFSS).

# 6 Conclusion

We formalized Context Distribution Restoration (CDR) as a computational social science methodology for restoring incomplete survey metadata under known P(M). CDR combines individual predictive evidence with population distribution constraints via a recoverability-adaptive transport mechanism within an optimal transport framework that balances individual evidence and population fidelity: Sinkhorn enforces margins, while the adaptive mechanism regulates the accuracy–fidelity trade-off. Context anchors supply transport costs; they are not a standalone representation contribution. On CHNS, NHANES, and BRFSS, OT under P(M) restores population-consistent assignments with limited accuracy trade-offs, and a CHNS case study illustrates continuum structure in restored urban–rural metadata. Together, these elements offer a principled restoration layer bridging sample-level prediction and population-level consistency in survey analysis workflows.

## Declarations

### Availability of data and materials

The CHNS data are available from the Carolina Population Center (https://www.cpc.unc.edu/projects/chns). NHANES and BRFSS data are publicly available from the CDC (https://www.cdc.gov/nchs/nhanes/ and https://www.cdc.gov/brfss/). Analysis code is available from the corresponding author upon reasonable request.

### Competing interests

The authors declare that they have no competing interests.

### Funding

*[To be completed by the authors.]*

**Authors' contributions**

*[To be completed by the authors.]*

**Acknowledgements**

*[To be completed by the authors.]*

# References

Adler NE, Rehkopf DH (2008) U.S. disparities in health: descriptions, causes, and mechanisms. Annu Rev Public Health 29:235–252

Braveman P, Egerter S, Williams DR (2011) The social determinants of health: coming of age. Annu Rev Public Health 32:381–398

Braveman PA, Cubbin C, Egerter S, Williams DR, Pamuk E (2010) Socioeconomic disparities in health in the United States: what the patterns tell us. Am J Public Health 100(S1):S186–S196

CDC (2023) Behavioral Risk Factor Surveillance System (BRFSS). Centers for Disease Control and Prevention. https://www.cdc.gov/brfss/

CDC/NCHS (2023) National Health and Nutrition Examination Survey (NHANES). https://www.cdc.gov/nchs/nhanes/

Chen T, Guestrin C (2016) XGBoost: A scalable tree boosting system. In: Proceedings of KDD, pp 785–794

Courtemanche C, Marton J, Ukert B, Yelowitz A, Zapata D (2017) Early impacts of the Affordable Care Act on health insurance coverage in Medicaid expansion and non-expansion states. J Policy Anal Manage 36(1):178–210

Cuturi M (2013) Sinkhorn distances: Lightspeed computation of optimal transport. In: Advances in Neural Information Processing Systems 26

Diez Roux AV, Mair C (2010) Neighborhoods and health. Ann N Y Acad Sci 1186:125–145

Dong X, Morales AJ, Jahani E, Moro E, Lepri B, Bozkaya B, Sarraute C, Bar-Yam Y, Pentland A (2020) Segregated interactions in urban and online space. EPJ Data Sci 9:20. https://doi.org/10.1140/epjds/s13688-020-00238-7

Duncan C, Jones K, Moon G (1998) Context, composition and heterogeneity: using multilevel models in health research. Soc Sci Med 46(1):97–117

Frogner C, Zhang C, Mobahi H, Araya-Polo M, Poggio T (2015) Learning with a Wasserstein loss. In: Advances in Neural Information Processing Systems 28

Geneletti S, Richardson S, Best N (2009) Adjusting for selection bias in retrospective, case–control studies. Biostatistics 10(1):17–31

Gozzi N, Comini N, Perra N (2024) Bridging the digital divide: mapping Internet connectivity evolution, inequalities, and resilience in six Brazilian cities. EPJ Data Sci 13:69. https://doi.org/10.1140/epjds/s13688-024-00508-8

Gretton A, Borgwardt KM, Rasch MJ, Schölkopf B, Smola A (2012) A kernel two-sample test. J Mach Learn Res 13:723–773

Groves RM, Peytcheva E (2008) The impact of nonresponse rates on nonresponse bias: a meta-analysis. Public Opin Q 72(2):167–189

Hagenaars JAJ, McCutcheon AL (eds) (2002) Applied Latent Class Analysis. Cambridge University Press, Cambridge

Kawachi I, Berkman LF (eds) (2003) Neighborhoods and Health. Oxford University Press, New York

Kashyap R, Verkroost FCJ (2021) Analysing global professional gender gaps using LinkedIn advertising data. EPJ Data Sci 10:39. https://doi.org/10.1140/epjds/s13688-021-00294-7

Kendall A, Gal Y (2017) What uncertainties do we need in Bayesian deep learning? In: Advances in Neural Information Processing Systems 30

King G, Honaker J, Joseph A, Scheve K (2001) Analyzing incomplete political science data: an alternative algorithm for multiple imputation. Am Polit Sci Rev 95(1):49–69

Krieger N (2001) Theories for social epidemiology in the 21st century: an ecosocial perspective. Int J Epidemiol 30(4):668–677

Lakshminarayanan B, Pritzel A, Blundell C (2017) Simple and scalable predictive uncertainty estimation using deep ensembles. In: Advances in Neural Information Processing Systems 30

Lazer D, Pentland A, Adamic L, et al. (2009) Computational social science. Science 323(5915):721–723

Li J, Zhou P, Xiong C, Hoi SCH (2020) Prototypical contrastive learning of unsupervised representations. arXiv:2005.04966

Link BG, Phelan J (1995) Social conditions as fundamental causes of disease. J Health Soc Behav Spec No:80–94

Little RJA, Rubin DB (2019) Statistical Analysis with Missing Data, 3rd edn. Wiley, Hoboken

Marmot M (2005) Social determinants of health inequalities. Lancet 365(9464):1099–1104

Marmot M, Wilkinson RG (eds) (2005) Social Determinants of Health, 2nd edn. Oxford University Press, Oxford

Mattsson C (2023) Computational social science with confidence. EPJ Data Sci 12. https://doi.org/10.1140/epjds/s13688-023-00435-0

Mokdad AH, Stroup DF, Giles WH (2003) Public health surveillance for behavioral risk factors in a changing environment. MMWR Recomm Rep 52(RR-9):1–12

Peyré G, Cuturi M (2019) Computational optimal transport. Found Trends Mach Learn 11(5–6):355–607

Phelan JC, Link BG, Tehranifar P (2010) Social conditions as fundamental causes of health inequalities: theory, evidence, and policy implications. J Health Soc Behav 51(S):S28–S40

Pickett KE, Pearl M (2001) Multilevel analyses of neighbourhood socioeconomic context and health outcomes: a critical review. J Epidemiol Community Health 55(2):111–122

Popkin BM, Du S, Zhai F, Zhang B (2010) Cohort Profile: The China Health and Nutrition Survey—monitoring and understanding socio-economic and health change in China, 1989–2011. Int J Epidemiol 39(6):1435–1440

Rubin DB (1987) Multiple Imputation for Nonresponse in Surveys. Wiley, New York

Sampson RJ, Raudenbush SW, Earls F (1997) Neighborhoods and violent crime: a multilevel study of collective efficacy. Science 277(5328):918–924

Särndal C-E, Swensson B, Wretman J (1992) Model Assisted Survey Sampling. Springer, New York

Sinkhorn R, Knopp P (1967) Concerning nonnegative matrices and doubly stochastic matrices. Pacific J Math 21(2):343–348

Snell J, Swersky K, Zemel R (2017) Prototypical networks for few-shot learning. In: Advances in Neural Information Processing Systems 30

Solar O, Irwin A (2010) A conceptual framework for action on the social determinants of health. Social Determinants of Health Discussion Paper 2. WHO, Geneva

van Buuren S (2018) Flexible Imputation of Missing Data, 2nd edn. CRC Press, Boca Raton

Villarroel MA, Blackwell DL, Jen A (2019) Tables of Summary Health Statistics for U.S. Adults: 2018 National Health Interview Survey. National Center for Health Statistics

Wang Y, Beydoun MA (2007) The obesity epidemic in the United States—gender, age, socioeconomic, racial/ethnic, and geographic characteristics: a systematic review and meta-regression analysis. Epidemiol Rev 29:6–28

WHO Commission on Social Determinants of Health (2008) Closing the Gap in a Generation. World Health Organization, Geneva

Williams DR, Collins C (2001) Racial residential segregation: a fundamental cause of racial disparities in health. Public Health Rep 116(5):404–416

Williams DR, Mohammed SA (2009) Discrimination and racial disparities in health: evidence and needed research. J Behav Med 32(1):20–47

Zhang B, Zhai FY, Du SF, Popkin BM (2014) The China Health and Nutrition Survey, 1989–2011. Obes Rev 15(S1):2–7

Zhou Y, Ma LJC (2000) Economic restructuring and suburbanization in China. Urban Geogr 21(3):205–236

Cuturi M, Doucet A (2014) Fast computation of Wasserstein barycenters. In: Proceedings of ICML, pp 685–693

Villani C (2009) Optimal Transport: Old and New. Springer, Berlin